\def\mode{0} 

\if 1\mode
    \documentclass[preprint,superscriptaddress,floatfix, nofootinbib]{revtex4-2}
    \usepackage{lineno}
    \linenumbers
    \usepackage{natbib}
\else
    \documentclass[twocolumn,superscriptaddress, aps, prx,floatfix, nofootinbib]{revtex4-2}
\fi

\usepackage{graphicx}
\usepackage{dcolumn}
\usepackage{bm}
\usepackage{float}
\usepackage{siunitx}
\usepackage{braket}
\usepackage{amsmath,amssymb}
\usepackage[draft]{hyperref}

\usepackage{ulem}
\begin{document}

\title{A Quantum Coherence Microscope in the Hubbard Regime}

\author{Lin Su}
\if 0\mode
    \email{ls4211@columbia.edu}
\fi
\affiliation{Department of Physics, Harvard University, Cambridge, Massachusetts 02138, USA}
\author{Michal Szurek}
\affiliation{Department of Physics, Harvard University, Cambridge, Massachusetts 02138, USA}
\author{Alec Douglas}
\affiliation{Department of Physics, Harvard University, Cambridge, Massachusetts 02138, USA}
\author{Ceren B.~Dag}
\affiliation{Department of Physics, Indiana University, Bloomington, Indiana 47405, USA}
\affiliation{Department of Physics, Harvard University, Cambridge, Massachusetts 02138, USA}

\author{Markus Greiner}
\if 0\mode
    \email{greiner@physics.harvard.edu}
\fi
\affiliation{Department of Physics, Harvard University, Cambridge, Massachusetts 02138, USA}

\date{\today}

\begin{abstract}
    Quantum coherence underlies collective quantum phenomena and emerging quantum technologies. Quantum gas microscopes have transformed quantum simulation by providing projective snapshots of many-body states with single-atom resolution, but spatially resolved measurements of off-diagonal correlations have remained elusive. Here, using the Talbot effect, we introduce a quantum coherence microscope that maps off-diagonal correlations onto site-resolved density signals with near-single-site resolution. We use this technique to locally probe the superfluid–Mott transition in a layer of a three-dimensional optical lattice and to measure coherence beyond nearest neighbors in an engineered potential landscape. By mapping off-diagonal correlations onto density signals through controlled Talbot evolution, this work opens new possibilities for accessing observables beyond the density basis through tailored matter-wave evolution and recapture.
\end{abstract}

\maketitle

\begin{figure*}
\includegraphics[width=\textwidth]{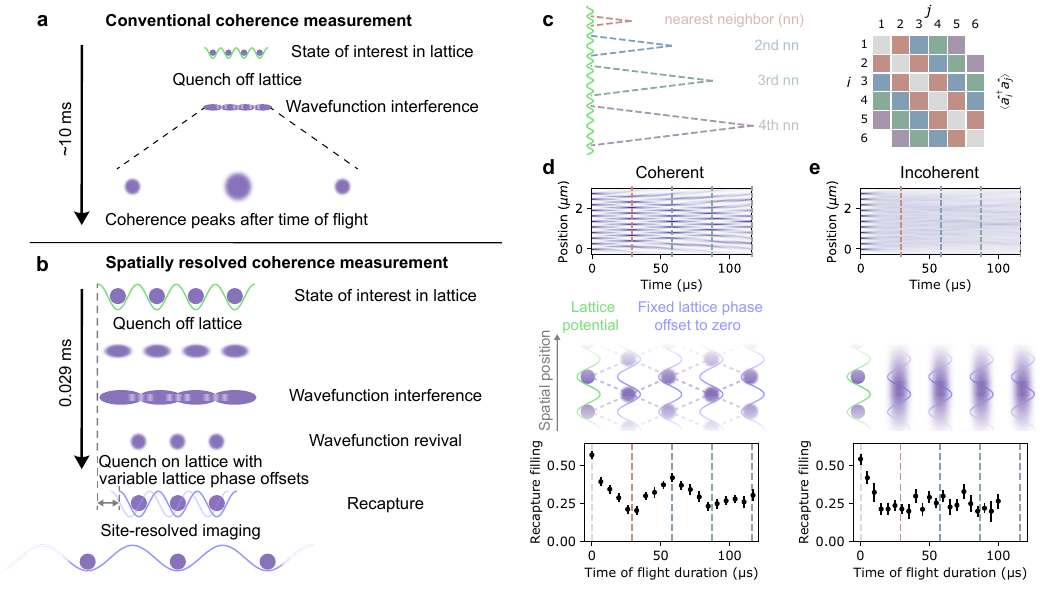}
\caption{\textbf{Short time of flight (ToF) with lattice recapture enables spatially resolved coherence measurements with wavefunction revivals.}
\textbf{a}, In a conventional probe of lattice coherence, the lattice (green) is abruptly turned off, and the atoms expand freely for $\sim 10~\mathrm{ms}$, producing interference peaks whose sharpness reflects the coherence averaged over the entire system.
\textbf{b}, To achieve spatial resolution, with the vertical lattice still on, the atoms undergo a short free-space ToF at integer multiples of half the Talbot time, $T_{\mathrm{T}}/2 = 29~\mu\mathrm{s}$, which is determined by the lattice spacing and atomic mass. The atoms are then recaptured in a phase-tunable lattice (blue) with the original (green) lattice spacing.
\textbf{c}, By varying the ToF, we can probe off-diagonal density-matrix elements corresponding to different lattice-site separations. $i$ and $j$ are site indices in a one-dimensional system.
\textbf{d}, For a phase-coherent array, varying the ToF reveals multiple revivals in the recaptured filling. Top, calculated time evolution of the perfectly coherent wave packets, showing revivals after specific times of flight. Colored dashed lines mark the predicted revival times associated with nearest neighbor, next nearest neighbor, and higher-order coherence. Middle, phase coherence between lattice sites leads to revivals, detected through lattice recapture. Bottom, experimentally measured recaptured filling as a function of ToF. The recapture filling does not start near unity likely due to imperfect quench transfer from the green to the blue lattice.
\textbf{e}, For a phase-incoherent array, realized in the regime where tunneling is much weaker than interactions and disorder, no oscillations are observed as the ToF is varied.
\label{fig: intro}}
\end{figure*}

\section*{Introduction}

Quantum coherence~\cite{baumgratz2014}—the ability of quantum particles to exist in superpositions while maintaining a relative phase—is central to quantum technologies~\cite{streltsov2017colloquium}, from computing~\cite{shor1999polynomial} to sensing~\cite{giovannetti2011advances}. It also underlies the emergence of collective phenomena in quantum many-body systems, such as superfluidity~\cite{penrose1956bose,yang1962concept}. The single-particle density matrix for a particle on a lattice can provide information beyond local density measurements: its diagonal elements describe occupations, whereas its off-diagonal elements encode coherence between particles at different lattice sites.

Ultracold atoms in optical lattices~\cite{Bloch2008} have become instrumental to the quantum simulation of strongly correlated matter. Combined with quantum gas microscopy~\cite{Bakr2009, Sherson2010}, they open the door to site-resolved studies of the Bose- and Fermi-Hubbard models with unprecedented microscopic detail~\cite{gross2021quantum,daley2022practical}, enabling measurements of correlation functions, nonlocal order parameters~\cite{endres2011observation}, and even entanglement entropy~\cite{islam2015measuring}. While local density distributions are now routinely measured, obtaining spatially resolved information about off-diagonal correlations remains elusive.

Time of flight (ToF) imaging is widely used to probe coherence in optical lattices~\cite{andrews1997observation,Greiner2002,Roth2003,gerbier2005phase,folling2005spatial,chin2006evidence,hadzibabic2006berezinskii,Mun2007,gericke2008high,trotzky2010suppression,braun2015emergence,baier2016extended,bergschneider2019experimental}. As illustrated in Fig.\ref{fig: intro}\textbf{a}, interference peaks emerge after expansion when phase coherence extends beyond a few lattice sites. However, conventional ToF measurements integrate contributions from different regions of the cloud, thereby obscuring spatial variations of the many-body state. This limitation is particularly consequential in inhomogeneous systems, where spatially varying chemical potentials can produce coexisting phases separated by domain walls, obscuring or even mischaracterizing states. At the opposite spatial extreme, coherence in isolated few-site systems has been measured~\cite{schumm2005matter,folling2007direct,sebby2007preparing}, including through singlet-triplet interferometry~\cite{trotzky2010controlling,greif2013short,hart2015observation}. Several approaches have also been proposed for probing phase coherence locally within extended systems~\cite{knap2013probing,kessler2014single,pena2018measuring,murthy2019direct}. Recently, measurement of kinetic operators with single-bond resolution has been demonstrated~\cite{Impertro2024,impertro2025strongly}.

A different route is provided by the Talbot effect~\cite{Talbot1836,Wen2013}, in which an initially periodic wave pattern reconstructs itself after a characteristic propagation distance or free-evolution time. The effect has been widely observed in both atomic and optical systems~\cite{Besold1997,Schlosser2023, Clauser1994,Chapman1995,Sanz2007,Hollmer2019, Deng1999, Song2011}. Talbot-based techniques have further been applied to atoms in optical lattices~\cite{miyake2011bragg,mark2011demonstration,buchhold2011creating,santra2017measuring,Makhalov2019,asteria2021quantum}. Recent experiments have extended these techniques to measurements of phase coherence in Josephson-coupled systems with many atoms per site~\cite{bruggenjurgen2026phase}, and even to the single-atom-per-site regime~\cite{koehn2025quantum}. Nevertheless, spatially resolved access to off-diagonal elements of the single-particle density matrix across multiple lattice spacings has remained limited.

Here, building on Talbot-based coherence measurements, we introduce a quantum coherence microscope that resolves off-diagonal correlations with near-single-site resolution in a Bose-Hubbard quantum simulator (Fig.\ref{fig: intro}\textbf{b}). We suddenly quench the lattice, allow the wavefunction to evolve for a short time, and then recapture the atoms in a phase-tunable lattice before imaging. By combining short Talbot evolution with single-atom-resolved imaging, the microscope offers access to spatially resolved off-diagonal coherence. In particular, by choosing the Talbot evolution time, the recapture signal can be made sensitive to coherence between sites separated by $r=|i-j|$, with a corresponding spatial resolution of approximately $r$ lattice sites (Fig.~\ref{fig: intro}\textbf{c}). Thus, for nearest-neighbor coherence ($r=1$), the measurement reaches single-site spatial resolution, while measurements at larger separations trade spatial resolution for access to longer-range coherence, allowing spatially coexisting phases to be distinguished within a quantum gas microscope.  Coherent initial states exhibit Talbot revivals that generate modulations in the recaptured filling (Fig.\ref{fig: intro}\textbf{d}), whereas states with random relative phases show no revivals and correspondingly no modulation (Fig.\ref{fig: intro}\textbf{e}).

\section*{Model}

Our quantum simulator consists of up to a few hundred atoms confined to a single layer of a three-dimensional optical lattice, and is described by the effective Hamiltonian
\begin{equation}
\label{eq: Hamiltonian}
\begin{aligned}
H ={}& -t\sum_{\langle i,j \rangle}
(\hat a^\dagger_{i}\hat a_{j}+\mathrm{h.c.})
-\sum_i\mu_i\hat n_i \\
&+\frac{U}{2}\sum_i\hat n_i(\hat n_i-1)
+\sum_{i<j}
V_{i,j}\hat n_i\hat n_j.
\end{aligned}
\end{equation}
Here $V_{i,j}$ denotes the isotropically repulsive dipolar interaction between erbium atoms~\cite{chomaz2023dipolar}. The tunneling $t$ is controlled via the lattice depth, the on-site interaction $U$ is tuned primarily via magnetic Feshbach resonances, and $\mu_i$ represents the spatially varying chemical potential (Supplemental Material, SM).

\begin{figure*}
\includegraphics[width=\textwidth]{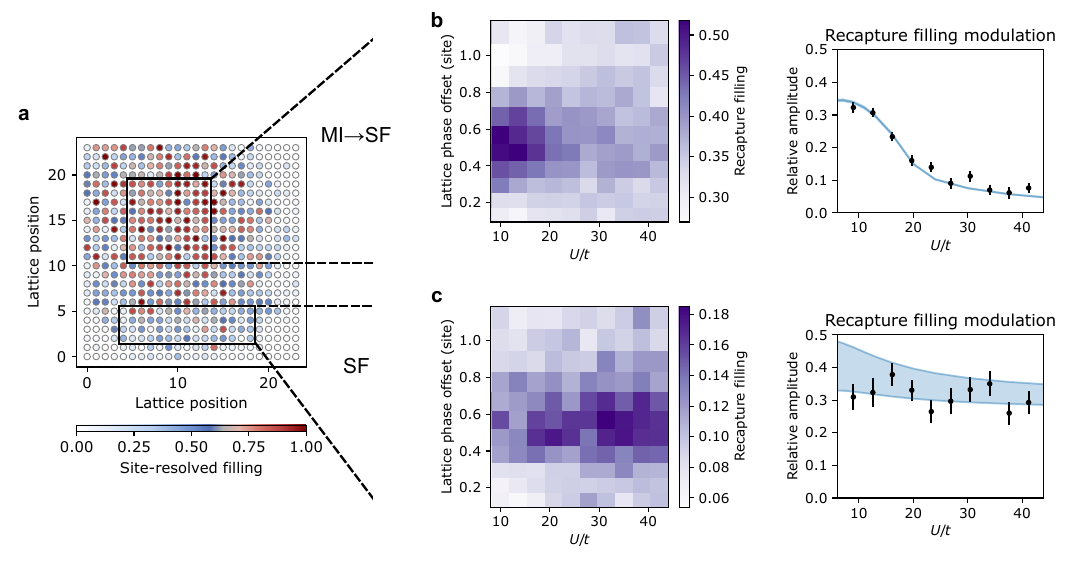}
\caption{\textbf{Spatially resolving the superfluid to Mott insulator transition in a harmonically confined atomic cloud.}
Without chemical potential engineering, a harmonically trapped atomic cloud in a lattice naturally exhibits different phases across the system. Using the quantum coherence microscope, we measure near-single-site resolved coherence.
\textbf{a}, With standard quantum gas microscopy, site-resolved images averaged over 77 shots show higher filling in the center and lower filling near the edges due to harmonic confinement with small $U/t$. We analyze two regions of interest: the center (roughly unit filling) and the edge (roughly one-third filling).
\textbf{b}, In the center region, the recapture filling modulation decreases as $U/t$ increases, signaling the superfluid to Mott insulator transition.
\textbf{c}, In the edge region, the modulation amplitude remains largely unchanged even at high $U/t$, indicating retained superfluid behavior. The recapture filling is lower than in the center because the initial density is lower.
Blue shaded areas show density matrix renormalization group simulations of the relative amplitude (with a global scaling parameter discussed in SM), and black points represent experimental measurements. Error bars denote the standard error of the mean.
\label{fig: Talbot_exp}}
\end{figure*}

We characterize the measured coherence using the normalized single-particle density matrix
\begin{equation}
\label{eq: coherence}
g^{(1)}_{i,j}
=
\frac{\langle \hat a^\dagger_{i}\hat a_{j}\rangle}
{\sqrt{\langle \hat n_{i}\rangle \langle \hat n_{j}\rangle}},
\end{equation}
where $i$ and $j$ denote lattice site indices and $\hat a_{i}$ and $\hat n_{i}$ are the annihilation and number operators, respectively. The Talbot evolution maps this off-diagonal coherence onto a modulation of the recaptured density as the phase of the recapture lattice is varied. At evolution times associated with a lattice separation $r$, the modulation is predominantly sensitive to $g^{(1)}_{i,i+r}$. Experimentally, we characterize this modulation by its relative amplitude $R$, which is related to the normalized coherence through $R=\eta g^{(1)}_{i,i+r}$, where $\eta$ accounts for the finite efficiency of the recapture protocol. Within the experimentally explored parameter regime, we find that $\eta$ is independent, within uncertainty, of lattice separation, filling, interaction strength, and Talbot evolution time, allowing a single global calibration to be applied across all measurements in this manuscript (SM).

\section*{Spatially resolving the superfluid to Mott insulator transition}

We first study the superfluid (SF) to Mott insulator (MI) transition~\cite{fisher1989boson} in a harmonically confined two-dimensional system. Harmonic confinement produces a spatially varying chemical potential in the form of $\mu_i\approx-k |\mathbf{i-i_0}|^2$, where $\mathbf{i_0}$ is the coordinate of the highest chemical potential site, and $k/t\approx0.17$ in our system. As shown in Fig.~\ref{fig: Talbot_exp}\textbf{a}, the central region is close to unit filling, whereas the outer regions remain substantially below unit filling. This inhomogeneity provides a natural benchmark for a spatially resolved coherence probe, because  different parts of the same cloud are expected to exhibit distinct coherence properties as interactions are increased.

For each value of $U/t$, we prepare the many-body state by tuning the on-site interaction $U$ using a magnetic Fano-Feshbach resonance and subsequently loading the atoms adiabatically into the optical lattice. Throughout this measurement, $U$ remains larger than the nearest neighbor repulsion $V$, while $t/V$ is maintained at sufficiently large values to stay away from the solid phases~\cite{Su2023}. We then perform Talbot imaging after half a Talbot time ($T_\mathrm{T}/2=29~\mu\mathrm{s}$) along both lattice directions, and recapture the atoms using a phase-tunable accordion lattice~\cite{Su2024}. By varying the recapture lattice phase (vertical axis in Fig.~\ref{fig: Talbot_exp}\textbf{b} and \textbf{c}), we observe a modulation in the recapture filling. We fit this modulation with a sinusoidal function, $A\cos(2\pi\phi)+O$, where $A$ is the amplitude, $O$ is the offset, and $\phi$ is the lattice phase offset measured in units of lattice sites. The relative modulation amplitude introduced above is then given by $R=|A|/O$.

\begin{figure}
\includegraphics[width=0.48\textwidth]{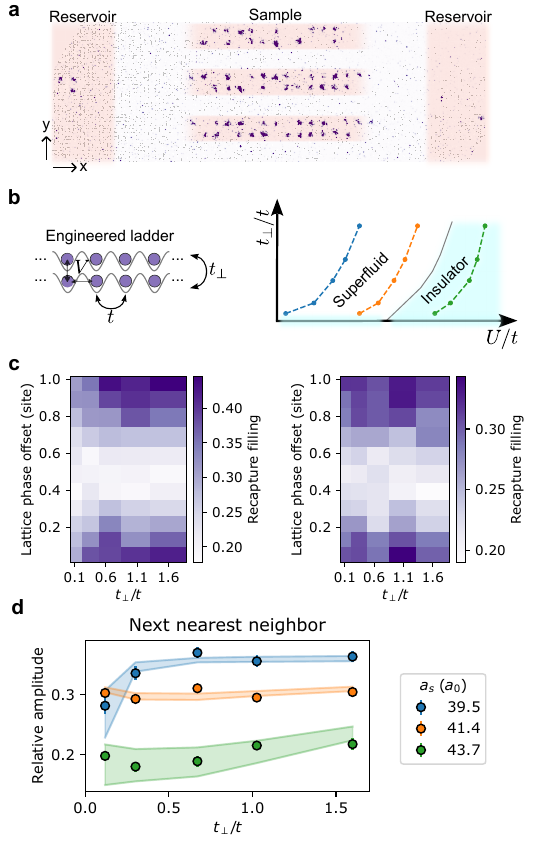}
\caption{\textbf{Probing coherence beyond nearest neighbors to explore the extended Bose-Hubbard model on a ladder geometry.}
\textbf{a}, In the experiment, we define distinct sample and reservoir regions, highlighted in red and overlaid on a single-shot image. We employ a quantum coherence microscope capable of spatially resolved coherence measurements to study the sample region only. 
\textbf{b}, We study a two-leg ladder system of soft-core bosons. As we couple the chains via tunneling $t_\perp$, the insulating phases, indicated by light-blue shading, are predicted to transition to a superfluid phase. The gray superfluid-insulator boundary is computed with density matrix renormalization group (SM). The three colored dashed lines indicate the positions where we took data.
\textbf{c}, We show the next nearest neighbor coherence along the chain in two regimes. The left (right) panels show results for the $U/t\approx 3$ ($12$) regime in our finite-size system with finite-rate state preparation along the blue (green) dashed line shown in panel \textbf{b}. In the low $U/t$ regime, the modulation of the recapture filling grows with $t_\perp$, signaling a rapid increase in coherence consistent with the superfluid regime, whereas it remains low in the Mott insulator regime.
\textbf{d}, We process the filling modulation to extract the relative coherence amplitude as a function of $t_\perp$ for different values of $a_s$. Shaded areas indicate time-evolving block decimation (TEBD) simulations implementing the ramps.
\label{fig: HI}}
\end{figure}

We observe that the central region exhibits a pronounced suppression of coherence as interactions are increased. As shown in Fig.~\ref{fig: Talbot_exp}\textbf{b}, the modulation amplitude decreases substantially as $U/t$ increases from 10 to 30, consistent with the loss of phase coherence expected across the superfluid to Mott insulator transition. In contrast, the edge region displays markedly different behavior. Despite increasing interactions, the modulation amplitude remains nearly the same up to $U/t\approx40$ (Fig.~\ref{fig: Talbot_exp}\textbf{c}), indicating that these lower density regions retain significant superfluid coherence. The experimental measurements are in excellent agreement with density matrix renormalization group (DMRG) calculations using a single global scaling parameter (SM).

\section*{Beyond nearest neighbor coherence in an engineered ladder}

Modern quantum simulators routinely use arbitrary potential shaping to redistribute entropy across the lattice for low-temperature state preparation~\cite{Bernier2009,chiu2018quantum,gross2021quantum,sompet2022realizing,xu2025neutral}. As illustrated in Fig.~\ref{fig: HI}\textbf{a}, a low-entropy sample region can be engineered and spatially separated from a surrounding reservoir. Combined with site-resolved imaging, such entropy engineering enables controlled studies of many-body systems within the sample region alone. Conventional ToF measurements average over the entire atomic cloud and therefore mix signals from both the sample and the reservoir, obscuring the properties of the low-entropy region of interest. In contrast, the quantum coherence microscope provides quantitative access to off-diagonal single-particle density matrix elements within the sample region only.

Our system consists of two one-dimensional chains coupled via tunneling ($t_\perp$) and isotropic density-density interactions ($V_{i,j}$) on approximately $12\times2$ lattice sites, realizing the extended soft-core Bose-Hubbard model. In the limit of vanishing interchain tunneling, the model hosts a rich phase diagram containing different insulating phases~\cite{Torre2006,Berg2008,Ashwath2024,Su2025Topological}. Our preparation is not perfectly adiabatic. We explore the system along the three colored dashed lines in Fig.~\ref{fig: HI}\textbf{b}, maintaining an isotropic nearest neighbor interaction $V\approx3t$ while tuning the on-site interaction via the effective s-wave scattering length $a_s$.

Because coherence in this ladder geometry is naturally defined along the chain direction, we measure off-diagonal single-particle density matrix elements only along the $x$ axis. To probe correlations beyond the nearest neighbor scale and better distinguish the different many-body states, we focus on the next nearest neighbor coherence, corresponding to a separation $r\equiv |i-j|=2$ sites in Eq.~\eqref{eq: coherence}.

For each parameter set, we vary the recapture phase and record the resulting filling modulation (Fig.~\ref{fig: HI}\textbf{c}). Distinct behaviors emerge for different regions of the parameter space. With $U/t\approx3$ along the blue dashed line in Fig.~\ref{fig: HI}\textbf{b}, the modulation amplitude increases immediately with finite interchain tunneling $t_\perp$, indicating a rapid increase in coherence consistent with the superfluid regime. In contrast, for larger $U/t\approx12$ along the green dashed line in Fig.~\ref{fig: HI}\textbf{b}, the modulation remains weaker throughout the explored parameter range, reflecting the tendency of the system to remain insulating even as $t_\perp$ increases.

From these filling modulations, we extract the relative next nearest neighbor modulation amplitude shown in Fig.~\ref{fig: HI}\textbf{d}. The blue trajectory exhibits the largest coherence, consistent with the strong coherence expected along this trajectory in the superfluid regime. In contrast, the green trajectory shows substantially weaker coherence throughout the sweep, in agreement with theoretical predictions that this parameter region does not enter the superfluid phase. We perform time-evolving block decimation simulations based on matrix product states that explicitly incorporate particle-number fluctuations, uncertainties in the scattering length, and non-adiabatic ramp dynamics. The resulting predictions, shown as shaded regions in Fig.~\ref{fig: HI}\textbf{d}, quantitatively reproduce the experimental observations using the same global scaling parameter employed throughout this work (SM).

These measurements demonstrate that the quantum coherence microscope can isolate and characterize coherence beyond nearest neighbors within a locally engineered region, without contributions from the surrounding reservoir. More broadly, the technique can be extended to probe coherence at even larger separations (SM), providing access to the spatial decay of off-diagonal correlations and enabling quantitative studies of quasi-long-range order and correlations exhibiting power-law decay in strongly interacting quantum matter~\cite{hadzibabic2006berezinskii}.

\section*{Conclusion and outlook}

In this work, we show that coherent matter-wave evolution followed by phase-controlled recapture maps off-diagonal elements of the single-particle density matrix onto site-resolved occupation signals, providing quantitative access to local coherence with near-single-site spatial resolution in the Hubbard regime. More broadly, our work establishes this combination of coherent evolution and recapture as a measurement protocol for quantum gas microscopy. Looking forward, this perspective opens the door to a broader class of protocols in which coherent transformations before projective imaging expand the measurement capabilities of atomic and molecular quantum systems. Such operations could be extended to fermionic systems~\cite{chin2006evidence,Murthy2014,murthy2015observation,brandstetter2025magnifying}, where programmable pre-imaging evolution has been proposed to reveal momentum distributions, Fermi surface structure, and even signatures of unconventional superfluidity~\cite{venu2023unitary,defossez2025dynamic,mark2025efficiently,weitenberg2026protocols,weitenberg2026protocols}.

\section*{Additional Information}

\paragraph*{Data and code availability}
The experimental data and analysis code supporting this study's findings are available from the corresponding authors upon request.

\paragraph*{Acknowledgements}
We are grateful for the contributions to building the experiment from O. Markovic, A. Krahn, A. Hebert, G. Phelps, R. Groth, S. F. Ozturk, S. Ebadi, S. Dickerson, and F. Ferlaino. We acknowledge fruitful discussions with R. Verresen and R. Sahay. 
We are supported by the U.S. Department of Energy Quantum Systems Accelerator DE-AC02-05CH11231, National Science Foundation Center for Ultracold Atoms PHY-1734011, Army Research Office Defense University Research Instrumentation Program W911NF2010104, Office of Naval Research Vannevar Bush Faculty Fellowship N00014-18-1-2863, Gordon and Betty Moore Foundation Grant GBMF11521, and Defense Advanced Research Projects Agency Optimization with Noisy Intermediate-Scale Quantum devices W911NF-20-1-0021. 

\paragraph*{Author contributions}
L.S., M.S., and A.D. contributed to building the experimental setup. L.S. conceived the idea of near-single-site resolved quantum coherence microscopy and performed the experiment and data analysis. L.S. and C.B.D. performed numerical simulations. L.S., C.B.D., M.G., and M.S. contributed to writing the manuscript. All authors discussed the results. M.G. and C.B.D. supervised the work.

\paragraph*{Competing interests}
M. Greiner is a co-founder, shareholder, and consultant of QuEra Computing. All other authors declare no competing interests.

\section*{Supplemental Material}

\subsection*{System parameters}

The Hamiltonian terms $t$, $U$, and $V$ are calibrated using methods described in~\cite{Su2025}. The effective s-wave scattering lengths~\cite{Patscheider2022, Su2025} are tuned via the magnitude of an external magnetic field and range from $40$ to $80\,a_0$, where $a_0$ is the Bohr radius. When probing the superfluid to Mott insulator transition in Fig.~\ref{fig: Talbot_exp}, tunneling is fixed at $t=h \times 6$ Hz ($h$ is the Planck constant), while the on-site interaction $U$ is tuned by varying the magnitude of the external magnetic field to achieve the desired values of $U/t$. When probing the engineered two-leg ladder system in Fig.~\ref{fig: HI}, tunneling is set to $t=h\times 8$ Hz, and the interchain tunneling $t_\perp$ is tuned by the lattice depth in the perpendicular direction. Since the Wannier function size changes with lattice depth, the on-site interaction $U$, which consists of contributions from effective $s$-wave scattering and dipolar interactions, is modified as $t_\perp$ is tuned. This effect results in the tilted dashed lines shown in Fig.~\ref{fig: HI}\textbf{b}. The magnetic dipole moment remains fixed at $7~\mu_B$~\cite{aikawa2012bose} throughout the experiment, where $\mu_B$ is the Bohr magneton. Throughout all experiments presented in this manuscript, the external magnetic field points perpendicular to the 2D plane, resulting in isotropically repulsive long-range interactions. Due to finite Wannier function sizes, the nearest neighbor repulsion is roughly $h\times23$ Hz and its decay with distance $r$ can be approximated by $V(r)=a/r^3 + b/r^4 + c/r^5 + d/r^6$, where $(a, b, c, d) = h \times (33.63, 0.13, -19.69, 9.08)$ Hz~\cite{Korbmacher2023}.

\subsection*{Connection to the optical Talbot effect}

\begin{figure*}
\includegraphics[width=\textwidth]{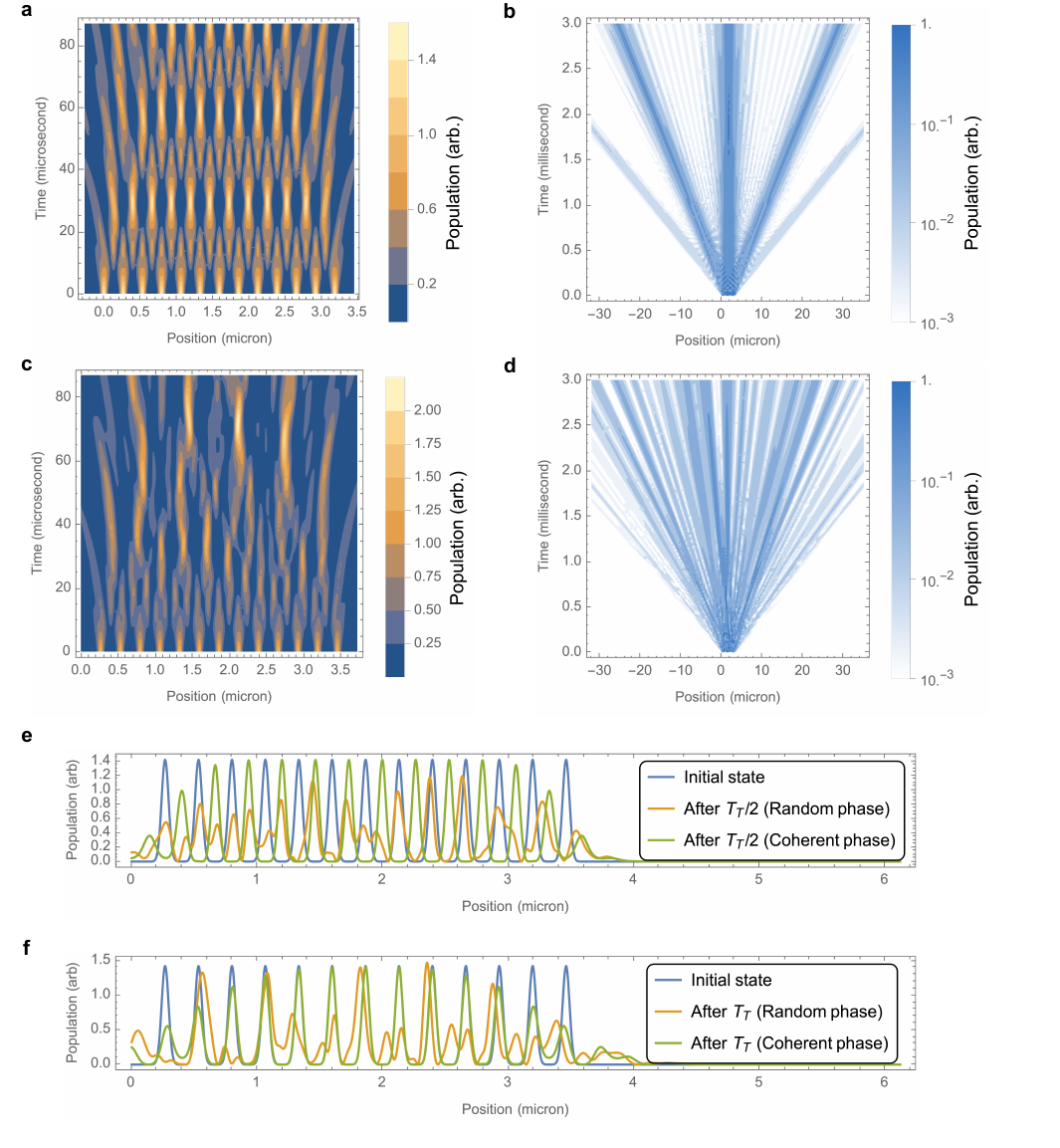}
\caption{\textbf{Talbot effect.} Matter-wave Talbot revivals complement ToF coherence-peak measurements by distinguishing states with coherent and incoherent phase relationships between neighboring sites. Time-evolution simulation results are shown here assuming a lattice depth of 25 recoil energies. Panels \textbf{a} and \textbf{b} assume all sites have the same phase, whereas panels \textbf{c} and \textbf{d} show one realization of independent random phases uniformly distributed across the lattice sites. Panel \textbf{e} shows horizontal cuts through panels \textbf{a} and \textbf{c} at times $t=0$ and $t=T_\mathrm{T}/2=29\mu\mathrm{s}$. Panel \textbf{f} shows horizontal cuts of \textbf{a} and \textbf{c} at times $t=0$ and $t=T_\mathrm{T}=58\mu\mathrm{s}$.
\label{fig: Talbot}}
\end{figure*}

The Talbot length in its original form is defined as $L=2d^2/\lambda$, where $d$ is the grating period and $\lambda$ is the de Broglie wavelength. The de Broglie wavelength can be expressed as $\lambda=h/p$, where the momentum is $p=\hbar k$. Thus, $L=2d^2\hbar k/h=d^2k/\pi$. On the other hand, the recoil energy is defined as $E_r=(h/(2d))^2/2m$. With $v=(\hbar k)/m$, we get $T_\mathrm{T}=mh/(2(h/(2d))^2)=h/(4E_R)$. For $^{164}$Er atoms in a lattice with 266 nm spacing, $T_\mathrm{T}\approx58$ microseconds.

We approximate the atomic wavefunction on each site of the optical lattice to be a Gaussian, which is a good approximation in deep lattices. In our experiment, the phases of interest occur at low tunneling and a lattice depth of approximately 25 recoil energies. Depending on the exact lattice depth, multiple orders of the fractional Talbot effect may show up. For example, at a lattice depth of 25 recoil energies, the time evolution of a phase coherent array of atoms in our optical lattice is shown in Fig.~\ref{fig: Talbot}\textbf{a}. At half of the Talbot time, with all sites coherent, the atomic wavefunction is concentrated right in between the original lattice sites. For a system of experimentally relevant size (12 sites are shown in the figure), the Talbot pattern revives many times. At much longer expansion times, orders of magnitude beyond the Talbot time, the momentum-space coherence peaks become spatially separated~\cite{Greiner2002}, as illustrated in Fig.~\ref{fig: Talbot}\textbf{b}. In contrast, when the atomic wavefunctions on different sites have uncorrelated phases, these wavefunction revivals and coherence peaks are not observed, as shown in Fig.~\ref{fig: Talbot}\textbf{c} and Fig.~\ref{fig: Talbot}\textbf{d}.

\subsection*{Numerical simulations}

The numerical results shown as shaded regions are obtained using the iTensor Julia package~\cite{itensor2022}.

In Fig.~\ref{fig: Talbot_exp}\textbf{b} (\ref{fig: Talbot_exp}\textbf{c}), we simulate the ground state of the Hamiltonian. The simulation system size is $x=6$ sites by $y=16\ (12)$ sites, with periodic boundary conditions along the 6-site direction and open boundary conditions along the 16 (12) site direction. The nearest neighbor coherence defined in Eq.~\eqref{eq: coherence} is calculated along both directions; the two results define the boundaries of the shaded region. The maximum bond dimension is set to 600. The center-region simulation in Fig.~\ref{fig: Talbot_exp}\textbf{b} is performed with uniform chemical potential with a system size of $x=6$ by $y=16$ sites. Observables are calculated from the central 8 sites along the y direction to avoid edge effects. The edge-region simulation in Fig.~\ref{fig: Talbot_exp}\textbf{c} is performed with the experimentally measured harmonic confinement chemical potential along the 12-site direction with a system size of $x=6$ by $y=12$ sites. Observables are calculated in the central 4 sites along the y direction, consistent with our choice of region of interest as shown in panel \textbf{a}.

\begin{figure}
\includegraphics[width=0.48\textwidth]{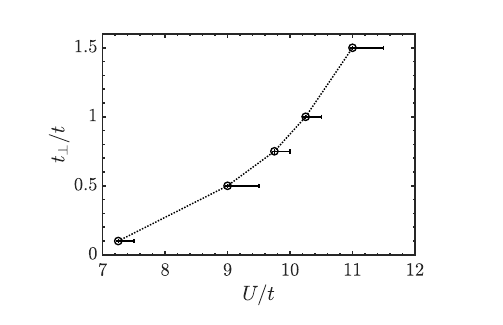}
\caption{\textbf{Phase transition between Mott insulator and superfluid in coupled 1D chains.} Large-system DMRG simulations identify the phase transition.
\label{fig: phaseDiagram}}
\end{figure}

Fig.~\ref{fig: HI}\textbf{b} shows a schematic of the phase diagram with the critical boundary between the Mott insulator and superfluid. Here we provide the calculated phase diagram as shown in Fig.~\ref{fig: phaseDiagram}, where the error bar stands for the uncertainty due to finite meshing of on-site interaction strength.         This phase diagram is obtained from both neutral- and charge-gap calculations using a DMRG algorithm on tensor networks. In these calculations, we used a maximum bond dimension of $\chi_m = 300$, an MPS compression truncation cutoff of $\epsilon \approx 10^{-10}$, and restricted the on-site occupation to $n = 0,1,2,3$ bosons. The dipolar interactions are included with the experimental imperfections up to $4^{\rm th}$ nearest neighbor along the chains, i.e.,~x-axis, and up to $2^{\rm nd}$ nearest neighbor on the diagonal direction. Both the neutral gap, i.e.,~$\Delta_0 = E_{2,N=L}-E_{1,N=L}$, and the charge gap $\Delta_c = E_{1,N=L+1}+E_{1,N=L-1}-2E_{1,N=L}$, where $E_{1,N=L+1}$ denotes the lowest energy of $L+1$ particles in a system of size $L$, vanish in the superfluid phase as $\propto L^{-1}$, whereas they do not vanish in the Mott insulator phase. Both gaps close at the same $U$ value for five different $t_{\perp}$ values, accurately determining the phase boundary.

In Fig.~\ref{fig: HI}\textbf{d}, we simulate the dynamical ramp of the experiment assuming we start from a superfluid ground state. Since the phase diagram is rich with multiple phases and critical points nearby, ground state simulation cannot capture our ramp results faithfully. The simulation system size is $x=16$ sites by $y=2$ sites with open boundary conditions in both directions. The edge chemical-potential profile produced by the repulsive walls projected through the objective is the same as that reported in~\cite{Su2025Topological} -- [0,1,3,10,32,85,191]~Hz. We assume we start with the ground state, and then follow the experimental ramp of $U$ and $t$ over the last 30 milliseconds of the experimental ramp. We checked that increasing the time duration of the ramp simulation does not affect the results. The time step is taken as 1.5 milliseconds per step, and the maximum bond dimension is set to 350. Increasing the bond dimension or decreasing the time step does not significantly affect the simulated results. Furthermore, we take into account the total atom number fluctuations by simulating 20, 22, and 24 particles in the system. Moreover, we take into account the uncertainty in on-site interaction $U$ of $\pm3$ Hz. All these simulations together result in the shaded area shown in Fig.~\ref{fig: HI}\textbf{d}.

\subsection*{Global scaling parameter}

\begin{figure}
\includegraphics[width=0.3\textwidth]{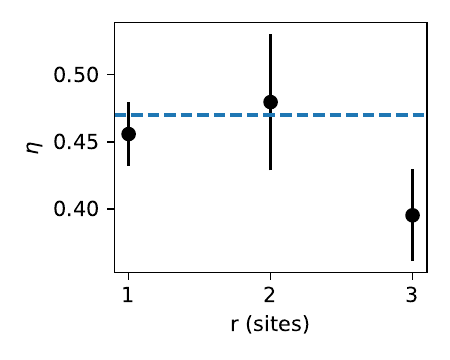}
\caption{\textbf{Consistency of the global scale factor $\eta$ across probed separations.}
Using the data presented in Fig.~1d, we examine the scale factor $\eta$ for different lattice separations $r$ probed by the coherence microscope. We use DMRG on a $6\times16$-site system to simulate the ground state and extract $g^{(1)}_{i,j}$. For each separation, $\eta$ is obtained from the ratio between the measured modulation amplitude and the simulated coherence. The values extracted for the three probed separations are statistically consistent with the global value $\eta=0.47$, shown by the blue dashed line.
\label{fig: dependence_of_eta}}
\end{figure}

To relate the experimentally measured relative recapture modulation $R$ to the normalized coherence $g^{(1)}_{i,j}$, we introduce a single global scale factor $\eta$, such that

$$R = \eta \cdot g^{(1)}_{i,j}.$$

We determine $\eta=0.47$ from a global calibration and use this same value for all theoretical curves shown in the manuscript.

Within our experimental uncertainties, the values of $\eta$ extracted under the different conditions explored in this work are consistent with this global value. In particular, we observe no statistically significant dependence on the probed separation $r$ (Fig.~\ref{fig: dependence_of_eta}). We similarly find no resolvable variation across the different ToF geometries (2D in Fig.~\ref{fig: Talbot_exp} or 1D in Fig.~\ref{fig: HI}), initial lattice depths, filling fractions, and $s$-wave scattering lengths used in the measurements presented.

\subsection*{Qualitative interpretation of the global scaling parameter}

We provide a qualitative explanation for why the global scaling parameter $\eta$ differs from unity. For the following discussion, we assume that the phases are perfectly coherent across lattice sites. Ideally, under the following conditions, the recapture filling should be zero: (1) the switching on and off of both lattices are instantaneous, (2) the phase of the blue recapture lattice is exactly 0.5 sites off, (3) the lattice depth of the green initial lattice is exactly the same as that of the recapture blue lattice, (4) there is no additional harmonic confinement, for example from the vertical lattice, and (5) interactions between atoms are small enough to be neglected. Under these conditions, based on exact diagonalization (ED) calculations, the energy of the final state per particle would be almost exactly the top of the recapture lattice potential, meaning that particles could escape the recapture lattice. However, in our experiment, these conditions are not exactly satisfied.

(1) Due to the finite rise time of the acousto-optic modulator and finite power feedback bandwidth, the fall time of the green lattice takes around 1 microsecond, and the rise time of the blue lattice takes around 2 microseconds. Our trap frequencies are roughly 50 kHz, so these rise and fall times are roughly an order of magnitude faster. However, the rise and fall times are still too long to be approximated as instantaneous. Based on ED, we estimate that the final particle energy shift due to the finite rise and fall times is 5\% of the final lattice depth.

(2) The blue recapture lattice is an accordion lattice and is projected via the high-numerical aperture objective. Because of the long free-space optical path of the accordion lattice (longer than 1 meter), shot-to-shot air current fluctuations can result in uncertainty of the lattice phase. We estimate a standard deviation of the phase to be 0.065 waves~\cite{Su2024}. These shot-to-shot phase fluctuations decrease the average energy of the recaptured particles and make them more likely to be trapped in the lattice. Based on ED, we estimate that the average final particle energy shift due to the shot-to-shot phase fluctuations of the blue lattice is 4\% of the final lattice depth.

(3) The blue accordion lattices that recapture the atoms are set to the maximum achievable power in our experiment to ensure high fidelity imaging later on. However, a slight mismatch in the Wannier function size of different lattice depths between the initial green lattice (from 20 to 30 recoil energies) and the recapture blue lattice (25 recoil energies) can result in slightly lower final particle energy. Based on ED, we estimate that the average final particle energy shift due to the lattice-depth mismatch between the green and blue lattices is 1\% of the final lattice depth. This is a rather small difference, so in regimes probed in our experiment with different green lattice depths, we do not see a significant difference.

Considering the first three factors together, assuming periodic boundary conditions and ignoring interactions between particles, we estimate that the average final particle energy is 91\% of the final lattice depth, meaning that the particles therefore lack sufficient classical energy to escape the lattice's confinement.

(4) Our vertical lattice is red-detuned and has a depth of more than 10 kHz. Although this is almost an order of magnitude smaller than the in-plane lattices, the confinement can still limit the high-energy atoms from escaping the trap. The in-plane lattices are also red-detuned and can prevent the atoms from escaping freely. Later on, during imaging, we expand the spacing of the blue accordion lattices. During this time, we introduce further dynamics due to the differences in lattice depth at different sites from the harmonic confinement~\cite{Su2025}.

(5) Although the free-expansion dynamics occur rapidly (29 microseconds or multiples of this), the dipolar interaction between our magnetic atoms can actually reach an energy scale of tens of kilohertz at short interparticle separations (within 30 nanometers). Therefore, our assumption of non-interacting bosons may break down in certain cases. However, simulating multiple interacting particles in continuous space is computationally demanding, so we do not estimate the magnitude of this effect. In 1D, pioneering theoretical work~\cite{Philipp2019} showed that contact interactions reduce the Talbot contrast, a prediction later confirmed experimentally~\cite{Wei2024} using a molecular gas with an exceptionally large effective $s$-wave scattering length of up to $2000\,a_0$ and density of around 100 particles per cubic micron. In our experiment, we do not quench the magnetic field to reduce the effective s-wave scattering length to zero. We also do not have the capability to turn off the magnetic dipolar interactions between the erbium atoms. However, experimentally, we do not see a significant decrease of $\eta$ for different ToF durations (within error bars, we see the same $\eta$ for Fig.~\ref{fig: Talbot_exp}, Fig.~\ref{fig: HI}, and Fig.~\ref{fig: dependence_of_eta}), so we qualitatively estimate that this factor may not be very significant within the regimes probed. This may be partly due to the fact that most of our experiments are conducted at a relatively small effective $s$-wave scattering length of approximately $40\,a_0$, which is more than two orders of magnitude smaller than the value demonstrated in the molecular-gas experiment~\cite{Wei2024}. In addition, the density of our system is on the order of 50 particles per cubic micrometer, more than an order of magnitude lower than that in the 2D tube Bose-Einstein condensates demonstrated using quantum gas magnifiers~\cite{asteria2021quantum}.

These considerations identify several experimental mechanisms that reduce the recapture contrast from its ideal value and therefore provide a qualitative explanation for $\eta<1$. A quantitative prediction of $\eta$ would require modeling the complete interacting recapture dynamics and is beyond the scope of this work. Experimentally, however, we find that a single value of $\eta$ consistently describes all regimes explored here.

\bibliography{references}

\end{document}